\documentclass[reprint, superscriptaddress,  aps, pra]{revtex4-2}
\usepackage{amssymb}
\usepackage{amsmath}
\usepackage{graphics}
\usepackage{graphicx}
\usepackage{subfigure}
\usepackage{dcolumn}
\usepackage{bm}
\usepackage{hyperref}
\usepackage{color}
\usepackage{epstopdf}

\begin{document}

\title{The saddle-point exciton signature on high harmonic generation in 2D
hexagonal nanostructures}

\author{H.K. Avetissian}
\affiliation{Centre of Strong Fields Physics, Yerevan State University,
Yerevan 0025, Armenia}

\author{S.S. Israelyan}
\affiliation{Centre of Strong Fields Physics, Yerevan State University,
Yerevan 0025, Armenia}

\author{H.H. Matevosyan}
\affiliation{Institute of Radiophysics and Electronics NAS RA, Ashtarak 0203, Armenia}

\author{G.F. Mkrtchian}
\thanks{mkrtchian@ysu.am}
\affiliation{Centre of Strong Fields Physics, Yerevan State University,
Yerevan 0025, Armenia}

\begin{abstract}
The disclosure of basic nonlinear optical properties of graphene-like
nanostructures with correlated electron-hole nonlinear dynamics over a wide
range of frequencies and pump field intensities is of great importance for
both graphene fundamental physics and for expected novel applications of 2D
hexagonal nanostructures in extreme nonlinear optics. In the current paper,
the nonlinear interaction of 2D hexagonal nanostructures with the
bichromatic infrared driving field taking into account many-body Coulomb
interaction is investigated. Numerical investigation in the scope of the
Bloch equations within the Houston basis that take into account $e-e$ and $%
e-h$ interactions in the Hartree-Fock approximation reveals significant
excitonic effects in the high harmonic generation process in 2D hexagonal
nanostructures such as graphene and silicene. It is shown that due to the
correlated electron-hole nonlinear dynamics around the van Hove singularity,
spectral caustics in the high harmonic generation spectrum are induced near
the saddle point excitonic resonances.
\end{abstract}

\maketitle

\section{ Introduction}

Many of the optoelectronic properties of graphene \cite{Castro} and its
analog silicene \cite{Sil1,Sil2,Sil3} can be understood within a
non-interacting free charged carrier picture. The most pronounced feature of
these nanostructures is the characteristic linear dispersion relation of
massless Dirac fermions \cite{Novoselov} and the anomalous integer quantum
Hall effect \cite{Zhang}. The measured optical conductivity up to the
visible region is close to the value of $e^{2}/4\hbar $ \cite{Mak1}
predicted within the free-particle (FP) theory \cite{Stauber}. On the other
hand, since the screening length diverges at the charge neutrality point 
\cite{Castro}, one can expect the significant influence of the many-body
electronic interactions on the properties of hexagonal nanostructures.
Indeed, depending on the substrate material many-body electronic
interactions lead to departure from the linear dispersion relation \cite%
{Elias,Hwang,Park} and to the fractional quantum Hall effect \cite%
{Bolotin,Du0}. In graphene, the ratio of the Coulomb potential energy to the
kinetic one, that is the Wigner-Seitz radius, is independent of density. The
latter is defined as $r_{s}=e^{2}/(\hbar \mathrm{v}_{F}\epsilon )$, where $%
\epsilon $ is the background lattice dielectric constant of the system, $%
\mathrm{v}_{F}$ is the Fermi velocity. For intrinsic graphene $r_{s}\approx
2.4$ and since $r_{s}$ is also the "effective fine structure constant" for
graphene \cite{Elias}, this should result in considerable changes in
graphene's properties, including the opening of an energy gap \cite%
{Khveshchenko,Gorbar,Drut}. Experimental evidence for such phenomena is
absent. This discrepancy is resolved if one takes into account the screening
stemming from the valence electrons, which is almost $4$ for intrinsic
graphene \cite{Das}. The substrate-induced screening further suppresses
Coulomb interaction making graphene a weakly interacting system. For
example, the substrate SiO$_{2}$ reduces Wigner-Seitz radius to $%
r_{s}\approx 0.5$. At first glance this is a small value, however
electron-electron interaction can also significantly modify the linear
optical response of graphene-like materials due to excitonic effects \cite%
{Yang,Kravets,Mak,Jornada,Peres,Mishchenko,exc0,exc1,exc2}. These effects
are significant near Van Hove singularity (VHS) \cite{Van Hove} point in the
Brillouin zone (BZ) giving rise to a pronounced peak in the optical
absorption. Both the position and shape of this peak evidence the role of
strong Coulomb interactions \cite{Yang,Kravets,Mak,Jornada}. Instead of
simple free-free transitions, electron-hole correlated transitions take
place. These are revealed in the absorption spectrum of graphene in the
ultraviolet range. Note that for silicene the excitonic resonance is
expected in the visible range of spectrum.

The significance of many-body Coulomb interaction has also been shown for
ultrafast many-particle kinetics \cite{Knorr1,Knorr2,Knorr-book} and for the
perturbative nonlinear optics in graphene \cite{Sun-Wu,Sun,Cheng,Rostami}.
With the further increase of the pump wave intensity, one can enter into the
extreme nonlinear optical regime \cite{Avetissian-book}, where high order
harmonics generation (HHG) takes place. The HHG until the last decade has
been the prerogative of atomic systems. But with the advent of graphene and
other novel nanostructures, it becomes clear that HHG can be much more
efficient in these materials. There are several investigations devoted to
the HHG phenomenon in the monolayer \cite%
{Mikh-Ziegler,Mer,H4,Al-Naib,H7,H8,H10,Mer18,Zurr}, bilayer \cite{H3,Du,H16}%
, and gapped graphene \cite{H11,2019} nanostructures with the pump wave of
linear polarization. Since the observation of the HHG enhancement by the
elliptically polarized light in graphene by Yoshikawa et al. \cite{Yoshikawa}
, the polarization and optical anisotropy effects of HHG in graphene have
been attracting much interest \cite{Liu,H12,H13,H14,H15,Wang,Feng,Mer2022}
as it is distinct from the HHG in gases where HHG is significantly
suppressed with an increase of the ellipticity of a pump wave \cite{Budil}.
After successful adoption of three-step semiclassical model developed for
atomic HHG \cite{Lewenstein} to gapped nanostructures \cite{Vampa2014,Vampa}%
, there have been attempts to extend this model to graphene \cite%
{Zurr,Feng,2019-2,Mer2022}.In the three-step semiclassical model, at the
first step for the gapped system, there is a localization of the excited
electron-hole wave packet in the BZ around the minimum bandgap at the
instant of tunneling. For graphene, due to the vanishing bandgap depending
on the intensity, polarization, and frequency of the pump wave different
scenarios can occur. In particular, instead of the tunneling
ionization/excitation the resonant one photon or/and multiphoton excitation
of the Fermi-Dirac sea can take place \cite{Mer}, or the first step can be
initiated by non-diabatic crossing \cite{Zurr} of the valence band electron
trajectories through the Dirac points, where the transition dipole moment is
singular. For graphene, as well as for other nanostructures one should also
relax the condition for recombination \cite{Feng,Mer2022}. Due to the wave
packet spreading an annihilation at a relative electron-hole distance
comparable to lattice spacing, so-called imperfect recollision can take
place \cite{Gaarde}. With these modifications, one can explain the
enhancement of HHG yield in the elliptically polarized laser fields \cite%
{Feng} or in two-color laser fields at orthogonal polarizations \cite%
{Mer2022}.

Compared with the gaseous system there is also one important factor that can
significantly modify the three-step semiclassical model. As has been shown
in Ref. \cite{Uzan-Ivanov}, at VHS spectral caustics are induced resulting
in a strong amplification of the HHG signal. On the other hand, in
graphen-like nanostructures near VHS the many-body Coulomb interaction is
expected to be significant. Hence, it is of interest to clear up the
signature of electron-electron interaction on the extreme nonlinear optical
response of graphene-like nanostructures in the situation when the charged
carriers are accelerated up to the $M$\ saddle point in the BZ. The
importance of Coulomb interaction for HHG in graphene has been previously
predicted in Ref. \cite{Mer18}. The latter study was conducted near the
Dirac points where excitonic effects are weak.

In the present work, we investigate the influence of saddle-point excitons
on the HHG process in a 2D hexagonal nanostructure. The electron-electron
Coulomb interaction is taken into account in the scope of the Hartree-Fock
(HF) approximation applicable to the full BZ. This ansatz leads to a closed
set of integrodifferential Bloch equations for the single-particle density
matrix in the Houston basis. The carrier-carrier and carrier-phonon
scatterings are taken into account phenomenologically with the relaxation
term. As reference nanostructures, we consider graphene and silicene. For
the latter, we neglect the small gap due to the spin-orbit coupling, which
is irrelevant for the current study.

The paper is organized as follows. In Sec. II the model and the basic
equations are formulated. In Sec. III, we present the main results. Finally,
conclusions are given in Sec. IV.

\section{The model and a closed set of integro-differential equations}

We consider the interaction of a strong laser field, bichromatic or
monochromatic, with a two-dimensional hexagonal nanostructure such as
graphene and silicene. The electric field strength of the considering
wave-field can be written as: 
\begin{equation}
\mathbf{E}\left( t\right) =f\left( t\right) E_{0}\left( \hat{\mathbf{e}}\cos
\left( \omega _{0}t\right) +\hat{\mathbf{e}}^{\prime }\varepsilon \cos
\left( \omega _{0}^{\prime }t-\varphi \right) \right) ,  \label{field}
\end{equation}
where $f\left( t\right) =\sin ^{2}\left( \pi t/\tau \right) $\ is the
sin-squared envelope function, $\tau $\ is the pulse duration, $\hat{\mathbf{%
e}}$\ and $\hat{\mathbf{e}}^{\prime }$\ are unite polarization vectors in
the plane of 2D nanostructure ($XY$), $\omega _{0}$\ and $\omega
_{0}^{\prime }$\ are currier frequencies, $E_{0}$\ is the amplitude, $%
\varepsilon $\ and $\varphi $\ are the relative amplitude and phase of the
two waves, respectively. We take an eight-cycle fundamental laser field. In
the HF approximation we reduce the electron-electron Coulomb interaction
into the mean-field Hamiltonian \cite{Mer18}. As a result, we obtain a
closed set of equations for the interband polarization $\mathcal{P}(\mathbf{k%
},t)=\mathcal{P}^{\prime }(\mathbf{k},t)+i\mathcal{P}^{\prime \prime }(%
\mathbf{k},t)$ and for the distribution functions $\mathcal{N}_{c/v}\left( 
\mathbf{k},t\right) $ of the conduction/valence bands. Then, one can obtain
semiconductor Bloch equations in the HF approximation. We will consider the
latter in the Houston basis, i.e. the crystal momentum $\mathbf{k}$ is
transformed into a frame moving with the vector potential $\mathbf{k}_{0}=%
\mathbf{k}-\mathbf{A}$, where $\mathbf{A=-}\int_{0}^{t}\mathbf{E}\left(
t^{\prime }\right) dt^{\prime }$ is the vector potential and $\mathbf{E}$ is
the laser electric field strength. For compactness of equations atomic units
are used throughout the paper unless otherwise indicated. On the HF level
for an undoped system in equilibrium, the initial conditions $\mathcal{P}(%
\mathbf{k},0)=0$, $N_{c}(\mathbf{k},0)=0$, and $N_{v}(\mathbf{k},0)=1$ are
assumed, neglecting thermal occupations. In this case the equation for $%
N_{v}(\mathbf{k},t)$ is superficial. Thus, the Bloch equations with damping (%
$\Gamma $) within the Houston basis read 
\begin{equation*}
\partial _{t}\mathcal{N}_{c}(\mathbf{k}_{0},t)=-2\mathrm{Im}\left\{ \left[ 
\mathbf{E}\left( t\right) \mathbf{D}_{\mathrm{tr}}\left( \mathbf{k}_{0}+%
\mathbf{A}\right) \right. \right.
\end{equation*}%
\begin{equation}
\left. \left. +\Omega _{c}\left( \mathbf{k}_{0}+\mathbf{A},t;\mathcal{P},%
\mathcal{N}_{c}\right) \right] \mathcal{P}^{\ast }(\mathbf{k}_{0},t)\right\}
,  \label{1}
\end{equation}%
\begin{equation*}
\partial _{t}\mathcal{P}(\mathbf{k}_{0},t)=-i\left[ \mathcal{E}_{eh}\left( 
\mathbf{k}_{0}+\mathbf{A}\right) -i\Gamma \right] \mathcal{P}(\mathbf{k}%
_{0},t)
\end{equation*}%
\begin{equation*}
+i\left[ \mathbf{E}\left( t\right) \mathbf{D}_{\mathrm{tr}}\left( \mathbf{k}%
_{0}+\mathbf{A}\right) +\Omega _{c}\left( \mathbf{k}_{0}+\mathbf{A},t;%
\mathcal{P},\mathcal{N}_{c}\right) \right]
\end{equation*}%
\begin{equation}
\times \left[ 1-2\mathcal{N}_{c}(\mathbf{k}_{0},t)\right] ,
\label{2}
\end{equation}%
where 
\begin{equation}
\mathcal{E}_{eh}\left( \mathbf{k}\right) =2\mathcal{E}\left( \mathbf{k}%
\right) -\Xi _{c}(\mathbf{k},t;\mathcal{P},\mathcal{N}_{c})  \label{21}
\end{equation}%
is the electron-hole energy defined via the band energy 
\begin{equation}
\mathcal{E}\left( \mathbf{k}\right) =\gamma _{0}\left\vert f\left( \mathbf{k}%
\right) \right\vert ,  \label{3}
\end{equation}%
and many-body Coulomb interaction energy%
\begin{equation*}
\Xi _{c}(\mathbf{k},t;\mathcal{P},\mathcal{N}_{c})=\frac{2}{\left( 2\pi
\right) ^{2}}\int_{BZ}d\mathbf{k}^{\prime }V_{2D}\left( \mathbf{k-k}^{\prime
}\right)
\end{equation*}%
\begin{equation}
\times \left\{ f_{c}\left( \mathbf{k,k}^{\prime }\right) \mathcal{N}%
_{c}\left( \mathbf{k}^{\prime }\right) +f_{s}\left( \mathbf{k,k}^{\prime
}\right) \mathcal{P}^{\prime \prime }\left( \mathbf{k}^{\prime },t\right)
\right\} .  \label{4}
\end{equation}%
In Eqs. (\ref{3}) $\gamma _{0}$ is the transfer energy of the
nearest-neighbor hopping and the structure function is 
\begin{equation}
f\left( \mathbf{k}\right) =e^{i\frac{ak_{y}}{\sqrt{3}}}+2e^{-i\frac{ak_{y}}{2%
\sqrt{3}}}\cos \left( \frac{ak_{x}}{2}\right) ,  \label{5}
\end{equation}%
where $a$ is the lattice spacing. In Eq. (\ref{4}) 
\begin{eqnarray*}
f_{c}\left( \mathbf{k,k}^{\prime }\right) &=&\cos \left[ \mathrm{arg}f\left( 
\mathbf{k}^{\prime }\right) -\mathrm{arg}f\left( \mathbf{k}\right) \right] ,
\\
f_{s}\left( \mathbf{k,k}^{\prime }\right) &=&\sin \left[ \mathrm{arg}f\left( 
\mathbf{k}^{\prime }\right) -\mathrm{arg}f\left( \mathbf{k}\right) \right] .
\end{eqnarray*}%
The electron-electron interaction potential is modelled by screened Coulomb
potential \cite{Knorr2}: 
\begin{equation}
V_{2D}\left( \mathbf{q}\right) =\frac{2\pi }{\epsilon \epsilon _{\mathbf{q}%
}\left\vert \mathbf{q}\right\vert },  \label{6}
\end{equation}%
which accounts for the substrate-induced screening in the 2D nanostructure ($%
\epsilon $) and the screening stemming from valence electrons ($\epsilon _{%
\mathbf{q}}$). In Eqs. (\ref{1}) and (\ref{2}) the interband transitions are
defined via the transition dipole moment%
\begin{equation*}
\mathbf{D}_{\mathrm{tr}}\left( \mathbf{k}\right) =-\frac{a}{2\left\vert
f\left( \mathbf{k}\right) \right\vert ^{2}}\sin \left( \frac{\sqrt{3}}{2}%
ak_{y}\right) \sin \left( \frac{ak_{x}}{2}\right) \widehat{\mathbf{x}}
\end{equation*}%
\begin{equation}
+\frac{a}{2\sqrt{3}\left\vert f\left( \mathbf{k}\right) \right\vert ^{2}}%
\left( \cos \left( ak_{x}\right) -\cos \left( \frac{\sqrt{3}}{2}%
ak_{y}\right) \cos \left( \frac{ak_{x}}{2}\right) \right) \widehat{\mathbf{y}%
},  \label{7}
\end{equation}%
and the light-matter coupling via the internal dipole field of all generated
electron-hole excitations: 
\begin{equation*}
\Omega _{c}\left( \mathbf{k},t;\mathcal{P},\mathcal{N}_{c}\right) =\frac{1}{%
\left( 2\pi \right) ^{2}}\int_{BZ}d\mathbf{k}^{\prime }V_{2D}\left( \mathbf{%
k-k}^{\prime }\right)
\end{equation*}%
\begin{equation}
\times \left\{ \mathcal{P}^{\prime }\left( \mathbf{k}^{\prime },t\right)
+if_{c}\left( \mathbf{k,k}^{\prime }\right) \mathcal{P}^{\prime \prime
}\left( \mathbf{k}^{\prime }\right) -if_{s}\left( \mathbf{k,k}^{\prime
}\right) \mathcal{N}_{c}\left( \mathbf{k}^{\prime },t\right) \right\} .
\label{8}
\end{equation}%
\begin{figure}[tbp]
\includegraphics[width=.45\textwidth]{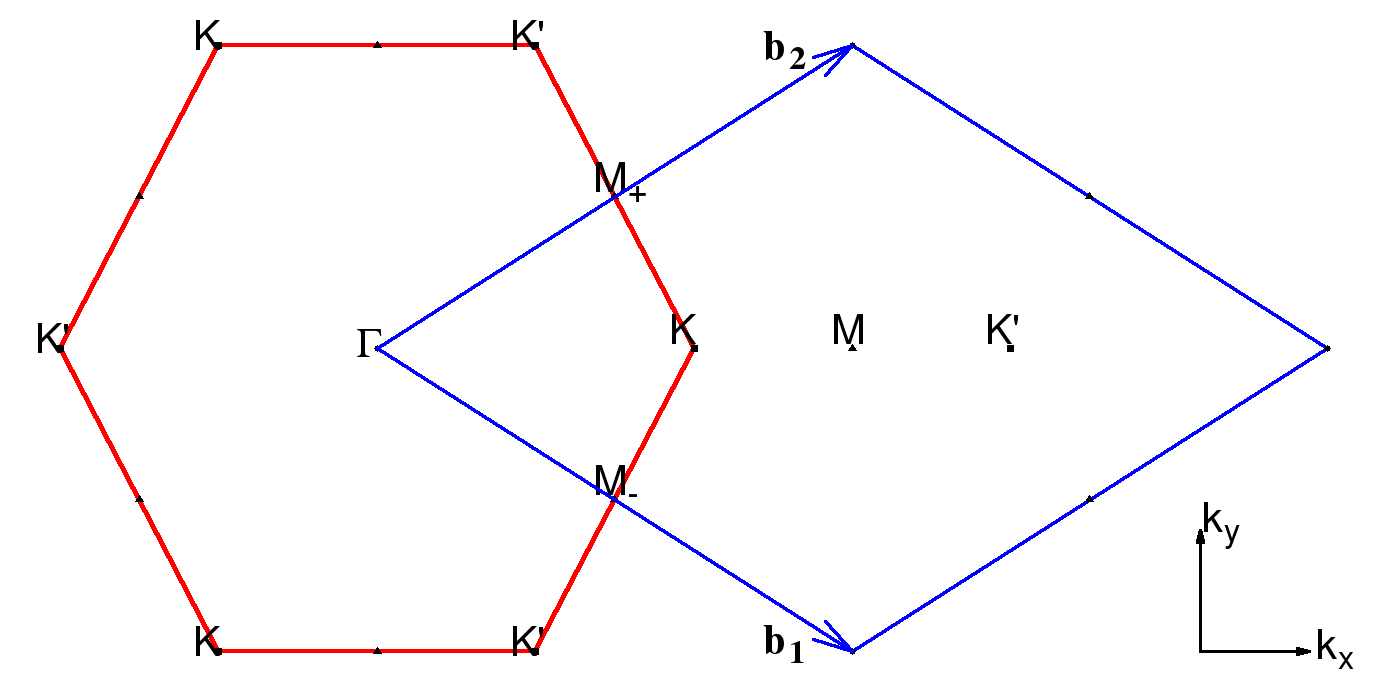}
\caption{The hexagonal first BZ of the reciprocal lattice with high-symmetry
points. The rhombus formed by the reciprocal lattice vectors is a reduction
of the second BZ and contains the same vectors of the first BZ.}
\end{figure}
The electromagnetic response in 2D hexagonal nanostructure is determined by
an intraband $\mathbf{j}_{a}\left( t\right) $ and interband $\mathbf{j}%
_{e}\left( t\right) $ contributions, which are given by 
\begin{equation}
\mathbf{j}_{a}\left( t\right) =-\frac{4}{(2\pi )^{2}}\int_{\widetilde{BZ}}d%
\mathbf{k}_{0}\left[ \mathbf{v}_{c}\left( \mathbf{k}_{0}+\mathbf{A}\right) 
\mathcal{N}_{c}\left( \mathbf{k}_{0},t\right) \right] ,  \label{ja}
\end{equation}%
\begin{equation}
\mathbf{j}_{e}\left( t\right) =-\frac{4}{(2\pi )^{2}}\int_{\widetilde{BZ}}d%
\mathbf{k}_{0}\mathrm{Re}\left[ \mathbf{v}_{\mathrm{tr}}^{\ast }\left( 
\mathbf{k}_{0}+\mathbf{A}\right) \mathcal{P}(\mathbf{k}_{0},t)\right] ,
\label{je}
\end{equation}%
respectively, where the band velocity is defined by $\mathbf{v}_{c}\left( 
\mathbf{k}\right) =\partial \mathcal{E}\left( \mathbf{k}\right) /\partial 
\mathbf{k}$, and $\mathbf{v}_{\mathrm{tr}}\left( \mathbf{k}\right) =2i%
\mathcal{E}\left( \mathbf{k}\right) \mathbf{D}_{\mathrm{tr}}\left( \mathbf{k}%
\right) \ $is the transition matrix element for velocity. The Brillouin zone
is also shifted to $\widetilde{BZ}=BZ-\mathbf{A}$.

The obtained Eqs. (\ref{1}) and (\ref{2}) formulate a closed set of
integro-differential equations. We will solve these equations numerically.
It is more convenient to make integration of these equations in the reduced
BZ which contains equivalent $k$-points of the first BZ, cf. Fig. 1\textrm{.}
The sampling $\mathbf{k}$-points are distributed homogeneously in the
reduced BZ according to Monkhorst and Pack mesh. For the convergence of the
results we take $500\times 500$ $\mathbf{k}$-points running parallel to the
reciprocal lattice vectors: $\mathbf{b}_{1}=\left( \sqrt{3}%
k_{b}/2,-k_{b}/2\right) $ and $\mathbf{b}_{2}=\left( \sqrt{3}%
k_{b}/2,k_{b}/2\right) $, where $k_{b}=4\pi /\sqrt{3}a$. In the reduced BZ
the low-energy excitations are centered around the two points $\mathrm{K}%
\left( k_{b}/\sqrt{3},0\right) $ and $\mathrm{K}^{\prime }\left( 2k_{b}/%
\sqrt{3},0\right) $. The saddle point is $\mathrm{M}\left( \sqrt{3}%
k_{b}/2,0\right) $. The time integration is performed with the standard
fourth-order Runge-Kutta algorithm. For sufficiently large 2D sample, when
generated fields are considerably smaller than the pump field $\left\vert 
\mathbf{E}^{(g)}\right\vert \ll \left\vert \mathbf{E}\right\vert $, the
generated electric field far from the hexagonal layer is proportional to the
surface current: $\mathbf{E}^{(g)}(t)=-2\pi \lbrack \mathbf{j}_{e}(t)+ 
\mathbf{j}_{a}(t)]/c$ \cite{Mer18}. For graphene/silicene on a substrate
with a refractive index of $n_{s}$, it is also necessary to take into
account the reflection of the incident wave \cite{Stauber} and rescale the
driving and the generated fields by a factor of $2/(1+n_{s})$. The HHG
spectral intensity is calculated from the fast Fourier transform of the
generated field $\mathbf{E}^{(g)}(\omega )$. For the substrate-induced
screening, we take $\epsilon =2.5$ that is close to the value of a graphene
layer on a SiO$_{2}$ substrate ($n_{s}\simeq 2$). The screening induced by
nanostructure valence electrons is calculated within the Lindhard
approximation of the dielectric function $\epsilon _{\mathbf{q}}$.

\section{Results}

The Coulomb contribution (\ref{4}) in Eq. (\ref{2}) describes the
renormalization of the single-particle energy $\mathcal{E}\left( \mathbf{k}%
\right) $ due to the repulsive electron-electron interaction. Note that in
the HF level we have neglected exchange interaction which is much smaller
compared to direct Coulomb term \cite{Hawrylak}. Since we consider an
undoped system, the exchange-correlation energy can also be neglected \cite%
{Polini}. Note that the Coulomb-induced constant self-energy has been
absorbed into the definition of the single-particle energy. At that, we will
fix the tight-binding parameter $\gamma _{0}$\ to obtain a good description
of high energies near VHS without loss of accuracy around the $K$ point. The
Coulomb contribution (\ref{8}) in Eq. (\ref{1}) accounts for electron-hole
attraction. This term gives rise to so-called saddle-point exciton \cite%
{Yang,Kravets,Mak,Jornada} near the VHS of hexagonal BZ. To validate our
theory within the limit of linear optics we first calculate the conductivity
for graphene ($\gamma _{0}\simeq 0.1\ \mathrm{a.u.}$ and $a=4.64\ \mathrm{%
a.u.}$) and for silicene ($\gamma _{0}=0.04\ \mathrm{a.u.}$ and $a=7.28\ 
\mathrm{a.u.}$). We assume the linearly polarized ($\hat{\mathbf{e}}=\left\{
1,0\right\} $, $\varepsilon =0$) laser with field strength $E\left( t\right)
=f\left( t\right) E_{0}\cos \left( \omega t\right) $. The conductivity can
be expressed as a function of the Fourier transform of the current density $%
\mathbf{j}\left( t\right) =\mathbf{j}_{a}\left( t\right) +\mathbf{j}%
_{e}\left( t\right) $ and the field strength:%
\begin{equation}
\sigma \left( \omega \right) =\frac{j_{x}\left( \omega \right) }{E\left(
\omega \right) }.  \label{sigma}
\end{equation}%
In Fig 2. we plot the FP and excitonic absorption spectrum via the real part
of the conduction $\sigma \left( \omega \right) $ versus laser field
frequency normalized to a universal one $\sigma _{0}=e^{2}/4\hbar $. From
this figure we see the characteristic redshifting in the excitonic
absorption spectrum with respect to the VHS peak expected in the scope of
the FP picture. We also see the asymmetric shape that arises from the
overlap with the free-particle transition. Due to ultrashort nature of the
driving pulse the maximum value and the widths of the peaks are somewhat
different than in the case of monochromatic wave \cite{Stauber}. 
\begin{figure}[tbp]
\includegraphics[width=.46\textwidth]{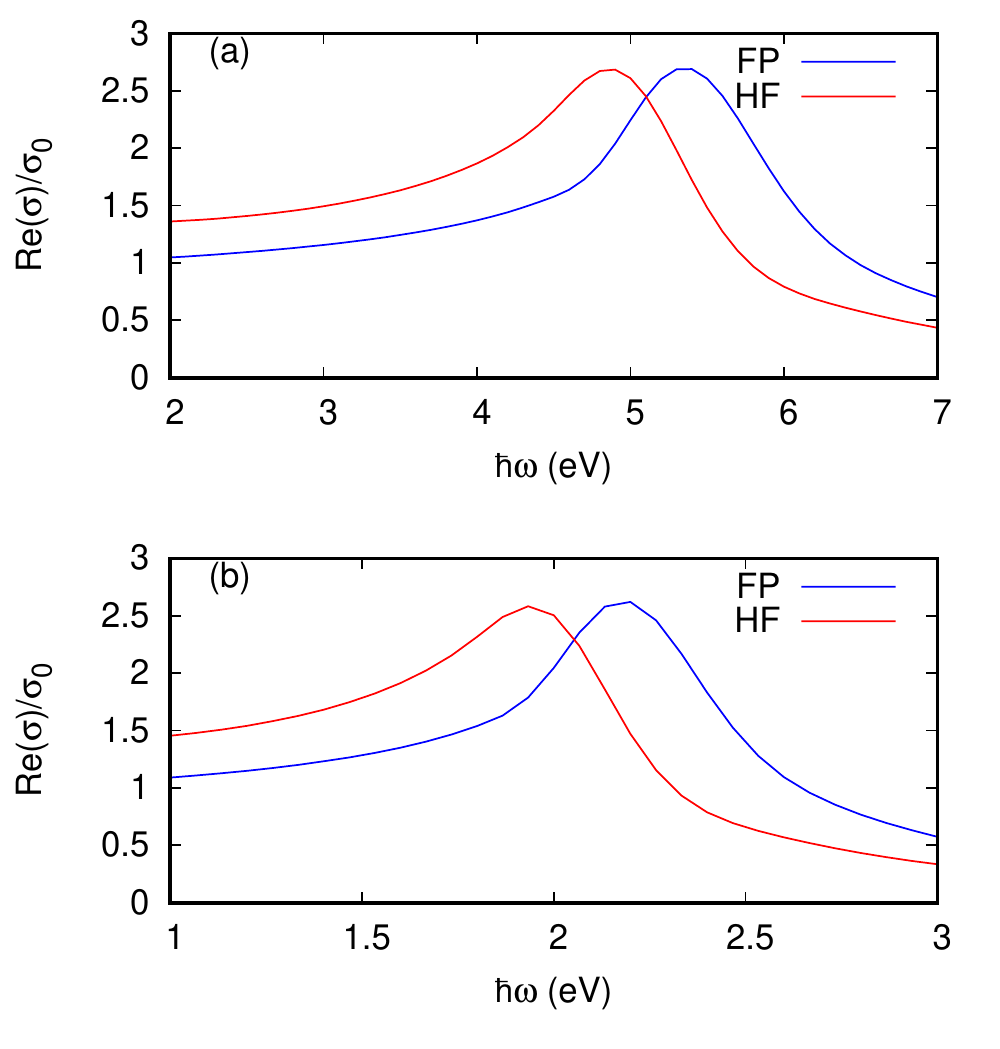}
\caption{Free-particle and excitonic absorption spectrum via the real part
of the conduction $\protect\sigma \left( \protect\omega \right) $ normalized
to universal one $\protect\sigma _{0}$ for graphene (a) and for silicene
(b). We take an eight-cycle laser field with the amplitude $E_{0}=10^{-5}\ 
\mathrm{a.u.}$ and the relaxation rate is taken to be $\Gamma =5\times
10^{-4}\ \mathrm{a.u.}$. The excitonic absorption spectrum is redshifted
with respect to the VHS peak expected in the scope of the free-particle
picture.}
\end{figure}
The significant changes in the absorption line shape and peak position near
the saddle point can be explained by the electron-hole interaction \cite%
{Yang}. The saddle-point excitonic resonances (SPER) have been extensively
investigated theoretically \cite{Phillips,Velicky,Kane,Balslev,SPE1}. The
changes in the absorption line shape can be understood from the band energy
near the $M$ -point. Near the saddle point $k_{M}=\left( \sqrt{3}%
k_{b}/2,0\right) $\ the band energy can be expanded as 
\begin{equation}
\mathcal{E}\left( \mathbf{k}_{M}+\delta \mathbf{k}\right) =\gamma _{0}+\frac{%
\delta k_{x}^{2}}{2m_{x}}+\frac{\delta k_{y}^{2}}{2m_{y}},  \label{near}
\end{equation}%
where $m_{x}=-2/\left( \gamma _{0}a^{2}\right) \ $and $m_{y}=2/\left(
3\gamma _{0}a^{2}\right) $. That is, along the $K-M$\ direction ($x$) the
effective mass is negative, while along the $\Gamma -M$\ direction ($y$) the
effective mass is positive. In Fig. 3, we show the band structure and
directions with effective masses of opposite signs. From the attractive
electron-hole interaction the development of quasi-discrete excitonic states
lying below the saddle-point singularity takes place, as is schematically
shown in Fig. 3. The exciton binding energy is the energy difference from
the SPER to the VHS calculated in the FP model. In our model, from Fig. 2,
we find binding energies of about $500\ \mathrm{meV}$\ and $250\ \mathrm{meV}
$\ for graphene and silicene, respectively. For graphene the obtained value
is close to experimental one \cite{Yang,Mak}. Although the electron-hole
interaction is attractive, the negative mass is equivalent to repulsion and
in the perpendicular direction these excitonic states do not lie below a
true gap. They consequently couple to the continuum formed by the band
descending from the saddle point. By this reason the overall absorption line
shape can be interpreted in terms of a Fano interference \cite%
{SPE1,Chae,Fano} effect. 
\begin{figure}[tbp]
\includegraphics[width=.42\textwidth]{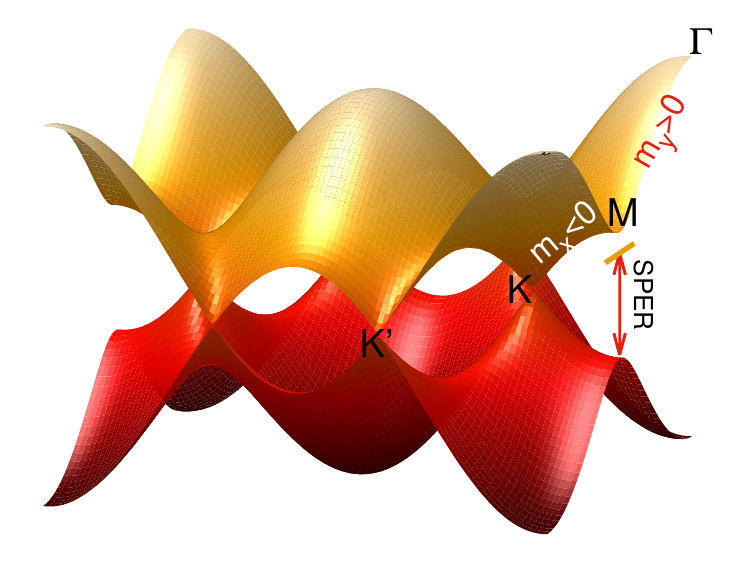}
\caption{The band structure of hexagonal nanostructure. The valence and
conduction bands cross at the $K$ and $K^{\prime }$ points. Near the saddle
point $M$ we show SPER and the band dispersion along the $K-M$ and $\Gamma -M
$ directions with effective masses of opposite signs.}
\end{figure}

Another feature of the excitonic resonance is the $\mathbf{k}$-space
redistribution \cite{Jornada} of the oscillator strength which is defined by
the interband polarization $\mathcal{P}(\mathbf{k},t)$. The excitonic states
are defined from Eq. (\ref{2}) when the pump field and relaxation rate are
set to zero: 
\begin{equation}
i\partial _{t}\mathcal{P}(\mathbf{k},t)=\mathcal{E}_{eh}\left( \mathbf{k}
\right) \mathcal{P}(\mathbf{k},t)-\Omega _{c}\left( \mathbf{k},t;\mathcal{P}
,0\right) .  \label{BS}
\end{equation}
This is the Bethe-Salpeter equation. For the excitonic states, the solution $%
\mathcal{P}(\mathbf{k},t)$ becomes more delocalized (localized) in the $%
\mathbf{k}$-space ($\mathbf{r}$-space) compared with the free particle
states. This effect along with coupling of excitonic states with the
continuum increases absorption below the SPER frequencies, cp. Fig. 2. These
effects can strongly affect the interband current (\ref{je}) also in
strong-field interaction regime. Hence, it is of interest to clear up the
signature of SPER on the extreme nonlinear optical response of the system,
when the generated harmonics' frequencies are near those resonances. In the
HHG process the frequency of the emitted harmonic is defined by the
electron-hole $\mathcal{E}_{eh}\left( \mathbf{k}_{0}+\mathbf{A}\right)$
energy which include kinetic energy acquired in the laser field, band gap
and also Coulomb interaction energy. That is, prior to electron-hole
annihilation their trajectories in the $\mathbf{k}$-space should be close to
the saddle point $M$.

Thus, to enhance excitonic effects there are two possibilities: to excite
the system with the photon of energy near $2\gamma _{0}$ \cite{2019-2} or
when photon energy is much smaller than $\gamma _{0}$ to accelerate
electrons/holes pair created near the Dirac points up to the energies $%
\gamma _{0}$. In the latter case, the trajectory in the $\mathbf{k}$-space
should pass close to the $M$ point along the positive mass direction. The
trajectory in the $\mathbf{k}$-space is the Lissajous diagram of the
corresponding vector potential $\mathbf{A}$. In the case of linear
polarization, this is impossible. In the case of circular/elliptic
polarization, the trajectory in the $\mathbf{k}$-space can be close to the $M
$\ point but being far from the Dirac point, which is the source of
electron-hole pairs \cite{Zurr}. For elliptic/circular polarization, the
initial electron-hole pairs can be produced by a resonant one photon or/and
multiphoton excitation of the Fermi-Dirac sea. However, for elliptic, and
especially for circular polarization of the pump wave, there is a shortage
of re-encountering electron-hole trajectories which is necessary for the
high-probability annihilation. For modest ellipticity $\varepsilon \simeq 0.3
$,\ the efficiency of the moderately high harmonics, where intraband current
(\ref{ja}) is significant, can be enhanced \cite{Yoshikawa}. However for
interband current, where SPER is expected, we need fields with $\varepsilon
\simeq 1$, since in the $\Gamma -M$\ direction we expect constructive
interference of many trajectories \cite{Uzan-Ivanov}, cp. Fig. 9. Thus, one
should choose a more sophisticated polarization of the driving waves. The
latter can be achieved via a bichromatic driving field that is composed of
the superposition of a fundamental pulse of linear polarization and its
harmonic at the orthogonal polarization. In the case of the second harmonic,
one may have "Infinity" sign-like figure in the $\mathbf{k}$-space that at
the sufficient intensity can pass through both $K$ and $M$ points. In this
case, it will approach the point $M$ in the direction $\Gamma -M$. Indeed,
this intuitive picture is validated with the numerical simulations: with the
mid-infrared pump pulses, we can see the fingerprint of the SPER on the high
harmonics. 
\begin{figure}[tbp]
\includegraphics[width=.5\textwidth]{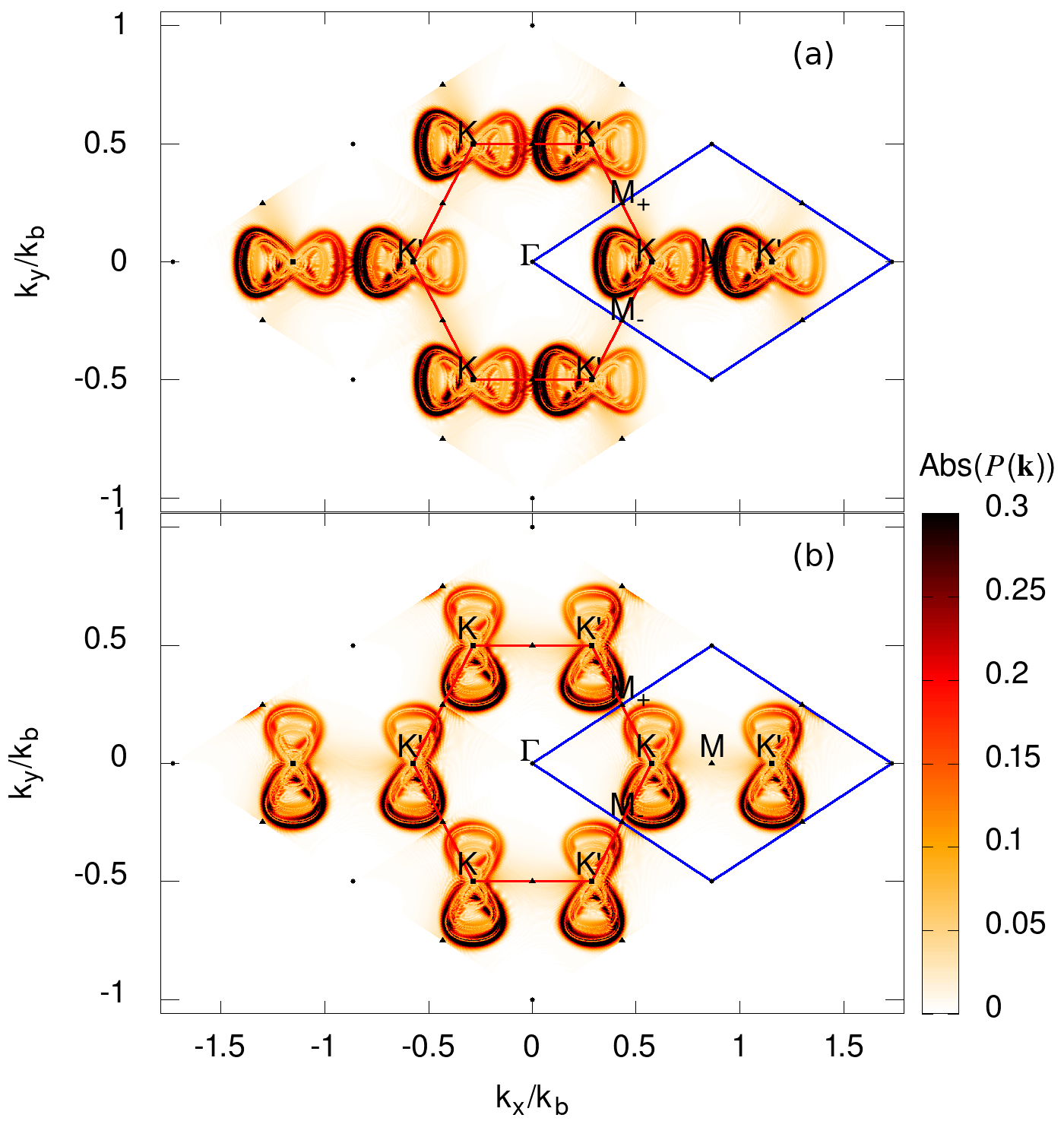}
\caption{The density plot of the absolute value of the interband
polarization (in arbitrary units) at the middle of interaction time ($t=%
\protect\tau /2$) in bichromatic laser field for $\hat{\mathbf{e}}=\left\{
1,0\right\} $, $\hat{\mathbf{e}}^{\prime }=\left\{ 0,1\right\} $ (a) and for 
$\hat{\mathbf{e}}=\left\{ 0,1\right\} $, $\hat{\mathbf{e}}^{\prime }=\left\{
1,0\right\} $ (b) as a function of scaled dimensionless momentum components (%
$k_{x}/k_{b}$, $k_{y}/k_{b}$). The fundamental frequency is $\protect\omega %
_{0}=0.1\ \mathrm{eV}/\hbar $, $\protect\omega _{0}^{\prime }=2\protect%
\omega _{0}$, $\protect\varepsilon =1$, $\protect\varphi =0$, and the
interaction parameter is $\protect\chi _{0}=1.7$. The first BZ (red hexagon)
with high symmetry points and reduced BZ (blue rhomb) are also shown.}
\end{figure}
\begin{figure}[tbp]
\includegraphics[width=.5\textwidth]{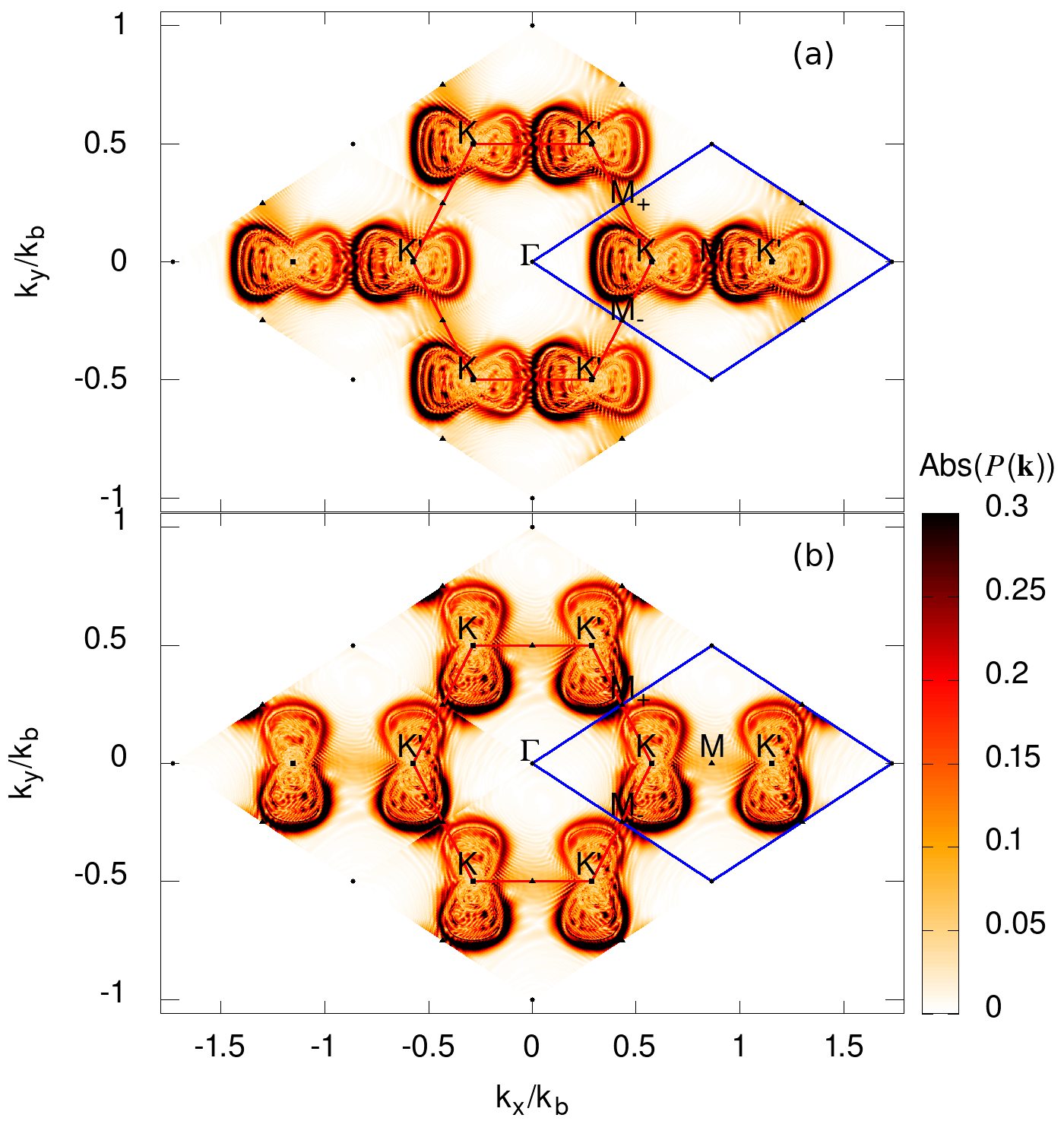}
\caption{The same as in Fig. (4), but for silicene.}
\end{figure}

The wave-particle interaction will be characterized by the dimensionless
parameter $\chi _{0}=eE_{0}a/\hbar \omega _{0}$ which represents the work of
the wave electric field $E_{0}$ on a lattice spacing in the units of photon
energy $\hbar \omega _{0}$. The parameter is written here in general units
for clarity. The total intensity of the laser beam expressed by $\chi _{0}$,
taking into account the reflectivity of the substrate, can be estimated as:%
\begin{equation*}
I_{\chi _{0}}=\chi _{0}^{2}\left( 1+\varepsilon ^{2}\right) \left(
1+n_{s}\right) ^{2}
\end{equation*}%
\begin{equation}
\times \lbrack \hbar \omega _{0}/\mathrm{eV}]^{2}\times \left[ \mathrm{%
\mathring{A}}/a\right] ^{2}\times 3.3\times 10^{12}\ \mathrm{W\ cm}^{-2}.
\label{int}
\end{equation}%
The amplitude ($A_{0}$) of vector potential can be expressed in terms of the
interaction parameter and reciprocal lattice spacing $k_{b}$ as $A_{0}=\chi
_{0}k_{b}\left( \sqrt{3}/4\pi \right) $. Thus, with an increase in $\chi _{0}
$ we can approach the point $M$ and thereby excite saddle-point excitons.
The parameter $\chi _{0}$ is varied up to $2$ and frequency up to $0.2\ 
\mathrm{eV}/\hbar $. Hence, the maximal intensity $\,\allowbreak 1.57\ 
\mathrm{TW/cm}^{2}$ impending on graphene is below the damage threshold \cite%
{Yoshikawa}. For silicene, due to larger lattice spacing the maximal
intensity is almost $2.5$ times smaller. In this paper, we consider a
two-band model formed from only $\pi $ orbitals. As a result, we neglect
transitions in $\sigma $\ and between $\pi -\sigma $\ orbitals. These
orbitals are separated from $\pi $\ orbitals by a large energy gap of $\sim
3\gamma _{0}$. Hence, we should restrict the pump wave field strength by the
condition $eE_{0}a<<3\gamma _{0}$, which is equivalent to $\chi
_{0}<<3\gamma _{0}/\hbar \omega _{0}$.

First we consider a bichromatic laser field with $\omega _{0}^{\prime
}=2\omega _{0}$, $\hat{\mathbf{e}}=\left\{ 1,0\right\} $, $\hat{\mathbf{e}}%
^{\prime }=\left\{ 0,1\right\} $, $\varphi =0$, and $\varepsilon =1$. At
these parameters the vector potential corresponding to the field (\ref{field}%
) draws $\infty -$like shape. In Figs. 4(a) and 5(a) the absolute value of
the interband polarization $\left\vert \mathcal{P}(\mathbf{k})\right\vert $
is shown at the middle of interaction time ($t=\tau /2$) for graphene and
silicene, respectively. It is clearly seen that the excitation patterns in
the Fermi-Dirac sea follow the Lissajous diagram of the vector potential. At
that, the surrounding of the $M$ point is excited in the $\Gamma -M$
direction and we expect the strong influence of this fact on the HHG
spectra. One can also consider the bichromatic crossed fields when the
polarizations are interchanged. In this case, we will have a vector
potential drawing eight-like shapes. As a result, the excitations of $M_{+}$%
\ and $M_{-}$saddle points will take place. These cases are shown in Figs.
4(b) and 5(b). 
\begin{figure}[tbp]
\includegraphics[width=.46\textwidth]{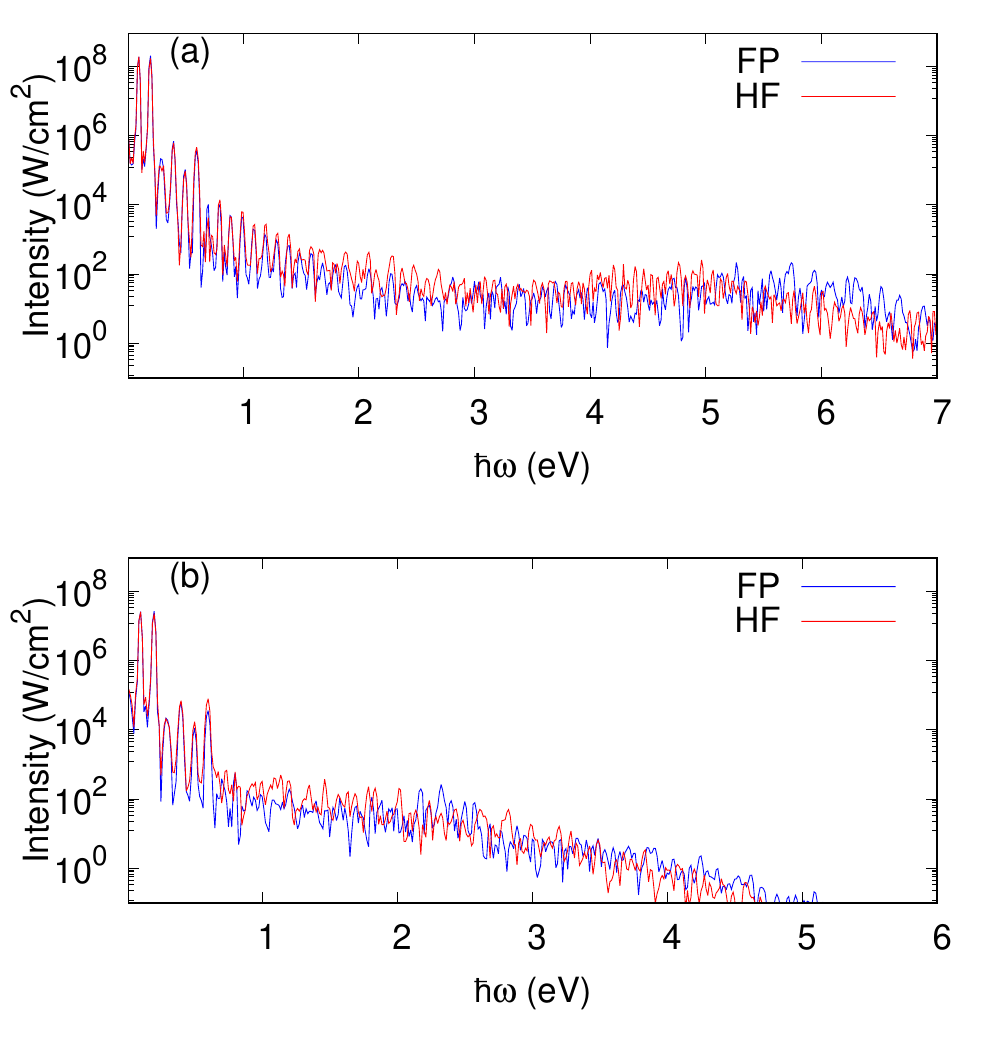}
\caption{The HHG spectra in logarithmic scale for graphene (a) and for
silicene (b) in the strong-field regime in the bichromatic driving field.
The fundamental frequency is $\protect\omega _{0}=0.1\ \mathrm{eV}/\hbar $
and the interaction parameter is $\protect\chi _{0}=1.7$. The relaxation
time is taken to be $\Gamma ^{-1}=2\protect\pi /\protect\omega _{0}\simeq
40\ \mathrm{fs.}$}
\end{figure}

In Fig. 6, the HHG spectra in logarithmic scale with $\omega +2\omega $
frequency mixing for graphene and silicene in the strong-field regime is
presented. We have also plotted the HHG spectra obtained in the scope of FP
model. As is seen from this figure, the intensity of high-harmonics are
enhanced near the frequencies close to SPER. For graphene, this is $4-5.0\ 
\mathrm{eV}$ and for silicene $1.5-2.0\ \mathrm{eV}$. The plateau peak is
redshifted compared with the free-particle case. For the beginning of the
spectrum where the intraband current (\ref{ja}) is dominant, the differences
with the free carrier picture are not so noticeable. We also need a
time-frequency analysis of the high harmonic spectrum for mapping the
harmonics near saddle-point excitonic resonances with the Lissajous diagram
of the vector potential. To this end, for graphene we perform the Morlet
transform ($\sigma =4\pi $) of the interband part of the surface current (%
\ref{je}): 
\begin{equation}
\mathbf{J}\left( t,\omega \right) =\sqrt{\frac{\omega }{\sigma }}%
\int_{0}^{\tau }dt^{\prime }\mathbf{j}_{e}\left( t^{\prime }\right)
e^{i\omega \left( t^{\prime }-t\right) }e^{-\frac{\omega ^{2}}{2\sigma ^{2}}%
\left( t^{\prime }-t\right) ^{2}}.  \label{wavelet}
\end{equation}%
The spectrogram, in a time interval where the waves' amplitudes are
considerable, is shown in Fig. 7 along with the Lissajous diagram of the
vector potential. The laser parameters correspond to Fig. 6(a). The numbers
over the spectrogram indicate the spectral caustics near the SPER. These
caustics take place with the period $0.5T$\ starting at $t\approx 3.75T$.
The corresponding points are shown on the Lissajous diagram of the vector
potential. As is clear from this mapping and also from Figs. 4 and 5, the
spectral caustics near SPER originate when electron-hole pair move in $%
\mathbf{k}$-space along $\Gamma -M$\ direction.\textrm{\ }Note that in this
direction the band velocity $v_{x}\left( \frac{2\pi }{a},k_{y}\right)
=\partial \mathcal{E}\left( \mathbf{k}\right) /\partial k_{x}=0$\
irrespective of $k_{y}$, the discrete states in $k_{y}$\ direction further
flatten the band near the saddle point making $v_{y}\simeq 0$. That is, near
the $\Gamma -M$\ direction the relative semi-classical velocity between the
electron and the hole vanishes, leading to a significant enhancement in
their annihilation rate \cite{Uzan-Ivanov}. 
\begin{figure}[tbp]
\includegraphics[width=.48\textwidth]{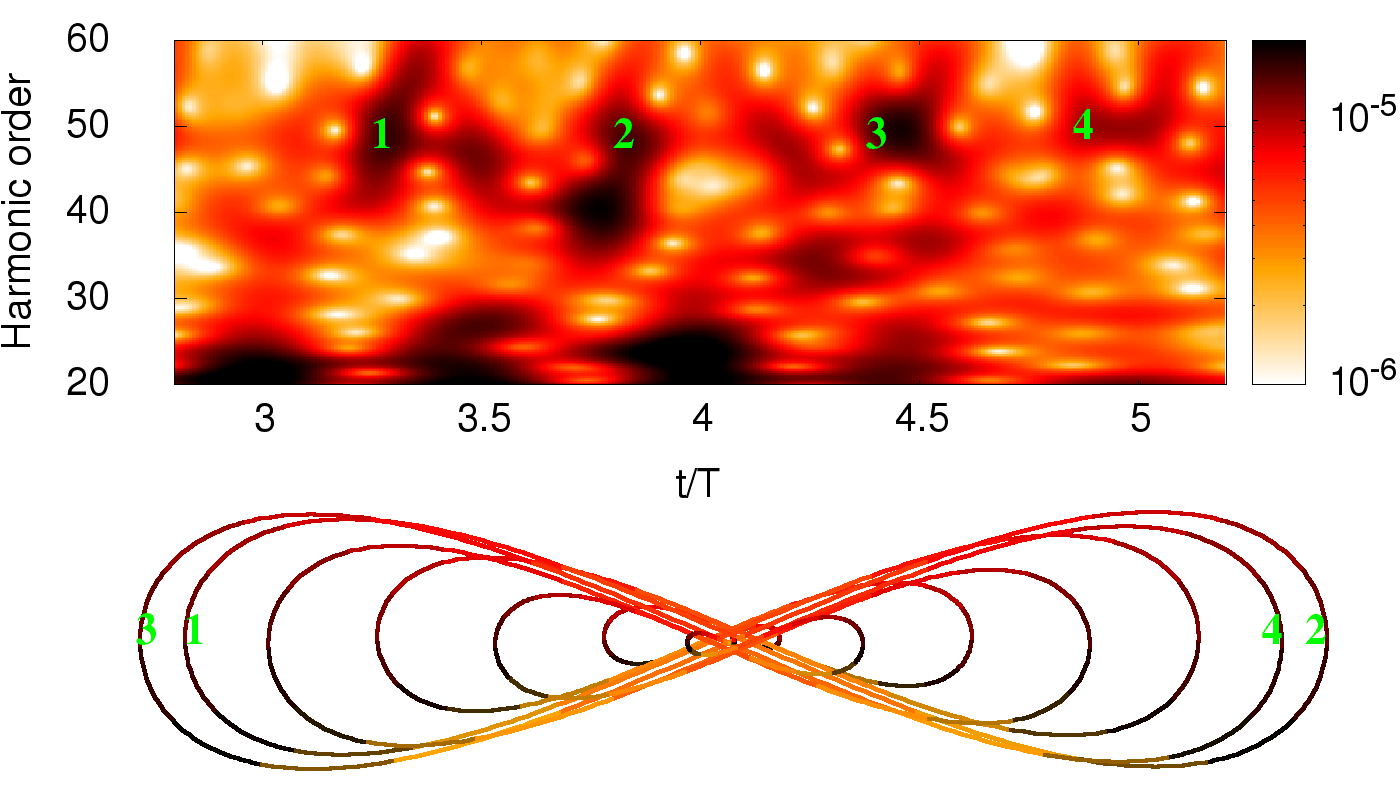}
\caption{The spectrogram (color box in arbitrary units)) of the HHG process
via the wavelet transform of the interband part of the surface current for
graphene. The lower panel shows the Lissajous diagram of the vector
potential. The laser parameters correspond to Fig. 6(a). The numbers over
the spectrogram and Lissajous diagram indicate the spectral caustics near
the SPER.}
\end{figure}
\begin{figure}[tbp]
\includegraphics[width=.49\textwidth]{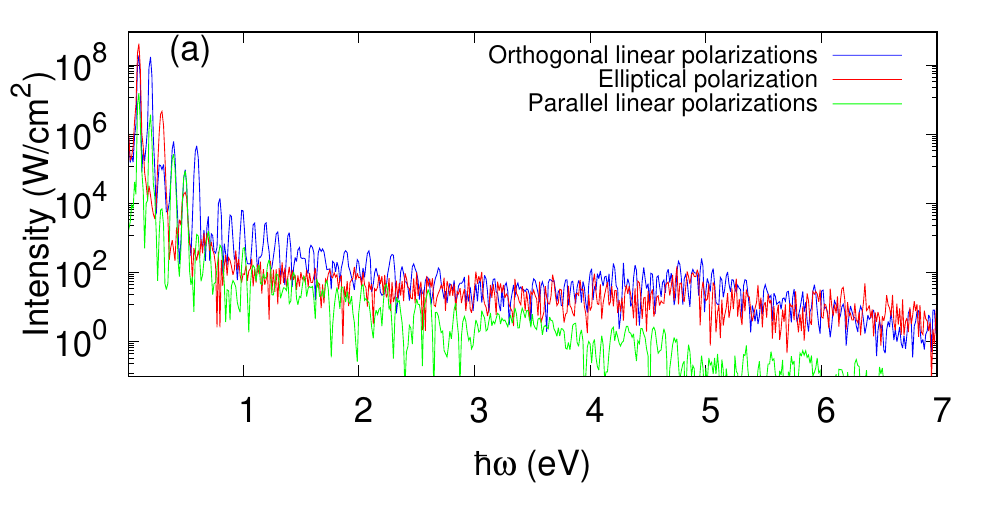}
\caption{The HHG spectra in logarithmic scale for graphene in the
strong-field regime for various polarizations of driving waves. The
relaxation time is taken to be $\Gamma ^{-1}=2\protect\pi /\protect\omega %
_{0}\simeq 40\ \mathrm{fs.}$ The fundamental frequency is $\protect\omega %
_{0}=0.1\ \mathrm{eV}/\hbar $. The intensities (\protect\ref{int}) for all
cases are equal. The interaction parameter for the orthogonal and parallel
polarization cases is $\protect\chi _{0}=1.7$. }
\end{figure}
To clarify the SPER signature in HHG spectra further, in particular with
respect to the polarization of driving waves, we also made numerical
calculations for the elliptical polarization of the laser field: $\omega
_{0}^{\prime }=\omega _{0}$, $\hat{\mathbf{e}}=\left\{ 1,0\right\} $, $\hat{%
\mathbf{e}}^{\prime }=\left\{ 0,1\right\} $, $\varphi =\pi /2$, $\varepsilon
=0.32$, and for the parallel linear polarizations of the bichromatic laser
field: $\omega _{0}^{\prime }=2\omega _{0}$, $\hat{\mathbf{e}}=\hat{\mathbf{e%
}}^{\prime }=\left\{ 1,0\right\} $, $\varphi =0$, and $\varepsilon =1$. The
results are shown in Fig. 8. For all three cases, we take the same intensity$%
\,$of \allowbreak $0.3\ TW/cm^{2}$. As is seen from this figure, in the case
of parallel linear polarization there is no enhancement near the SPER. In
the case of elliptic polarization, we have an enhancement in comparison to
linear polarization. However, the orthogonal polarization case is
preferable. For a qualitative understanding of this result, we also made a
semi-classical trajectory analysis taking into account the actual excitation
of the Fermi-Dirac sea, cp. Fig. 4. For the set of $\mathbf{k}_{0}$ points
in the region $\mathcal{E}_{eh}\left( \mathbf{k}_{0}\right) \leq 3\ \mathrm{%
eV}$ we integrated the equation $\mathbf{r}_{e}\left( t^{\prime },t\right)
=\int_{t^{\prime }}^{t}\left[ \mathbf{v}_{c}\left( \mathbf{k}_{0}+\mathbf{A}%
\left( t^{\prime \prime }\right) \right) \right] dt^{\prime \prime }$\ and
calculated the electron-hole distance $\rho \left( t^{\prime },t\right)
=\left\vert \mathbf{r}_{e}-\mathbf{r}_{h}\right\vert =2\left\vert \mathbf{r}%
_{e}\right\vert $. We kept only those trajectories for which at $t>t^{\prime
}$\ there is a local minimum of the electron-hole distance $\rho _{m}\left(
t^{\prime },t\right) <2a$, i.e. we have at list imperfect collision \cite%
{Gaarde}. Then we fixed the time and the corresponding energies $\mathcal{E}%
_{eh}\left( \mathbf{k}_{0}+\mathbf{A}\left( t\right) \right) $. In Fig. 9,
we plot the colliding trajectories that in the semiclassical picture
contribute to the caustic 2 indicated in Fig. 7 for the orthogonal and
elliptic polarization cases. As is seen from Fig. 9(a), in the orthogonal
polarization case the spectrum is dictated by the constructive interference
of many trajectories, while in the case of elliptic polarization, Fig. 9(b)
there is a shortage of re-encountering electron-hole trajectories.

\begin{figure}[tbp]
\includegraphics[width=.49\textwidth]{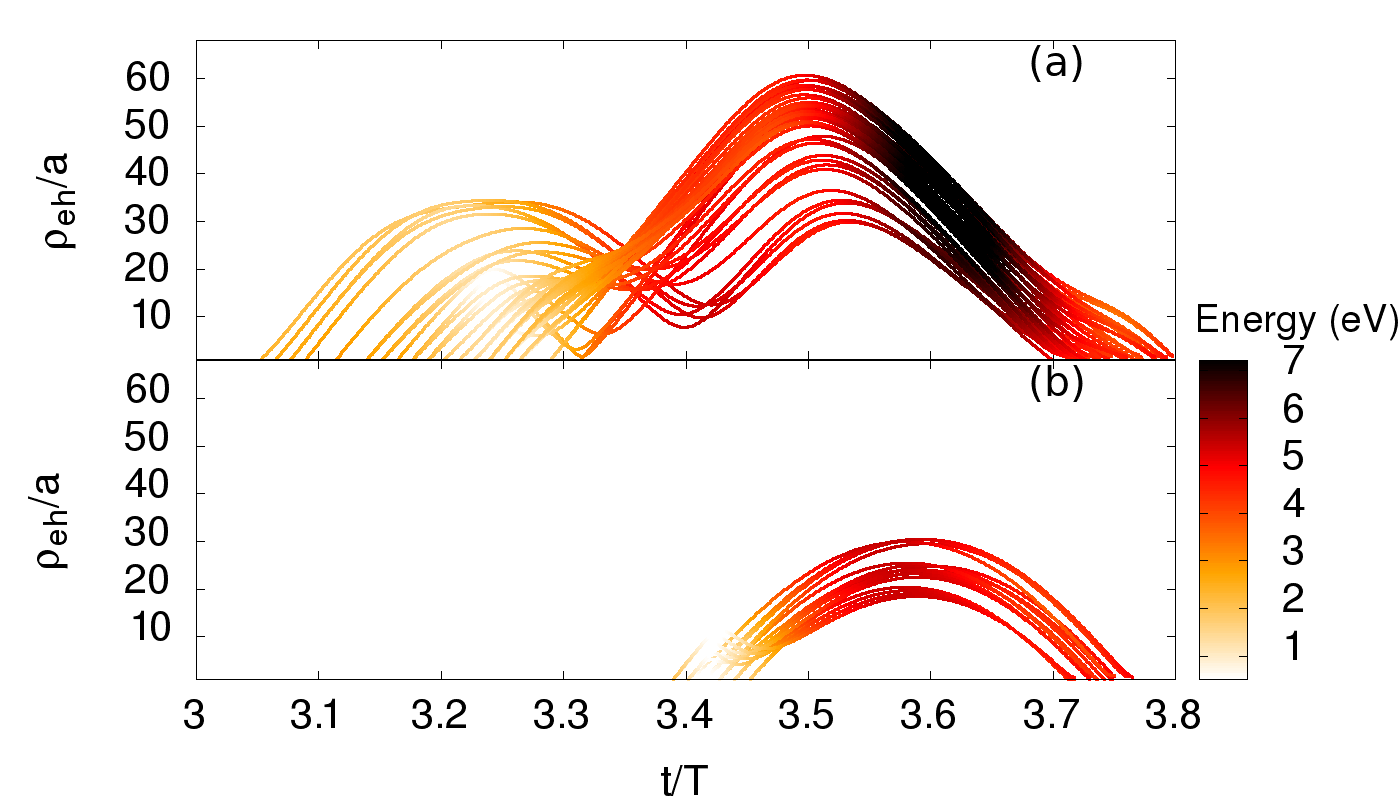}
\caption{Colliding trajectories. (a) The electron-hole distance as a
function of time for the trajectories which collide in the time interval
corresponding to the caustic 2 indicated in Fig. 7. The colored trajectory
and colored box show the energies acquired by carriers along the trajectory.
(b) The same as in (a) but for the elliptical polarization of the driving
wave with the parameters corresponding to Fig. 8.}
\end{figure}
\begin{figure}[tbp]
\includegraphics[width=.49\textwidth]{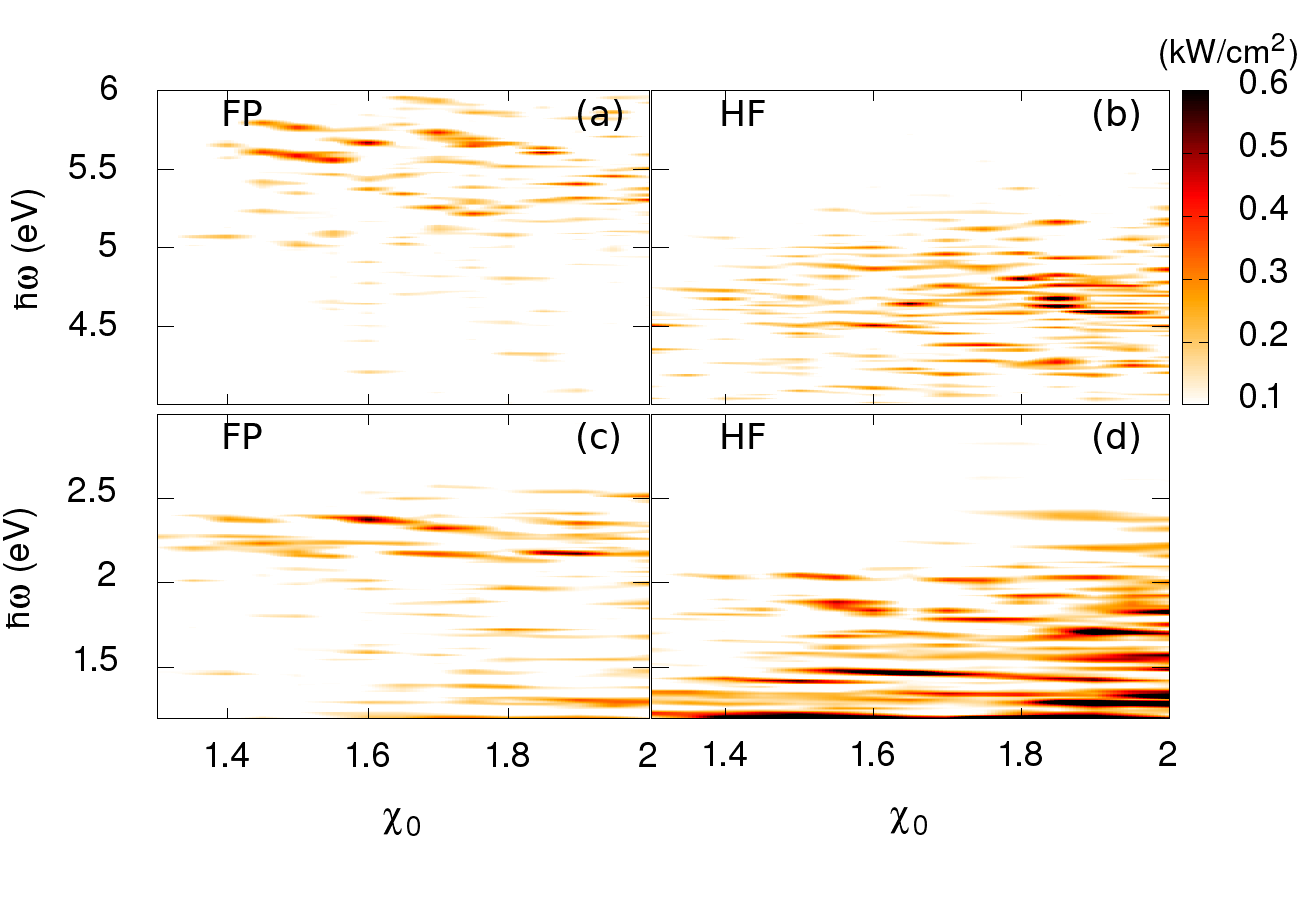}
\caption{The intensity of HHG as a function of the interaction parameter and
harmonic's photon energy for graphene (a,b) and silicene (c,d) in the
strong-field regime in the bichromatic driving field: $\protect\omega %
_{0}=0.1\ \mathrm{eV}/\hbar $, $\protect\omega _{0}^{\prime }=2\protect%
\omega _{0}$, $\hat{\mathbf{e}}=\left\{ 1,0\right\} $, $\hat{\mathbf{e}}%
^{\prime }=\left\{ 0,1\right\} $, $\protect\varphi =0$, and $\protect%
\varepsilon =1$ . The relaxation time is taken to be $\Gamma ^{-1}=40\ 
\mathrm{fs.}$}
\end{figure}
\begin{figure}[tbp]
\includegraphics[width=.49\textwidth]{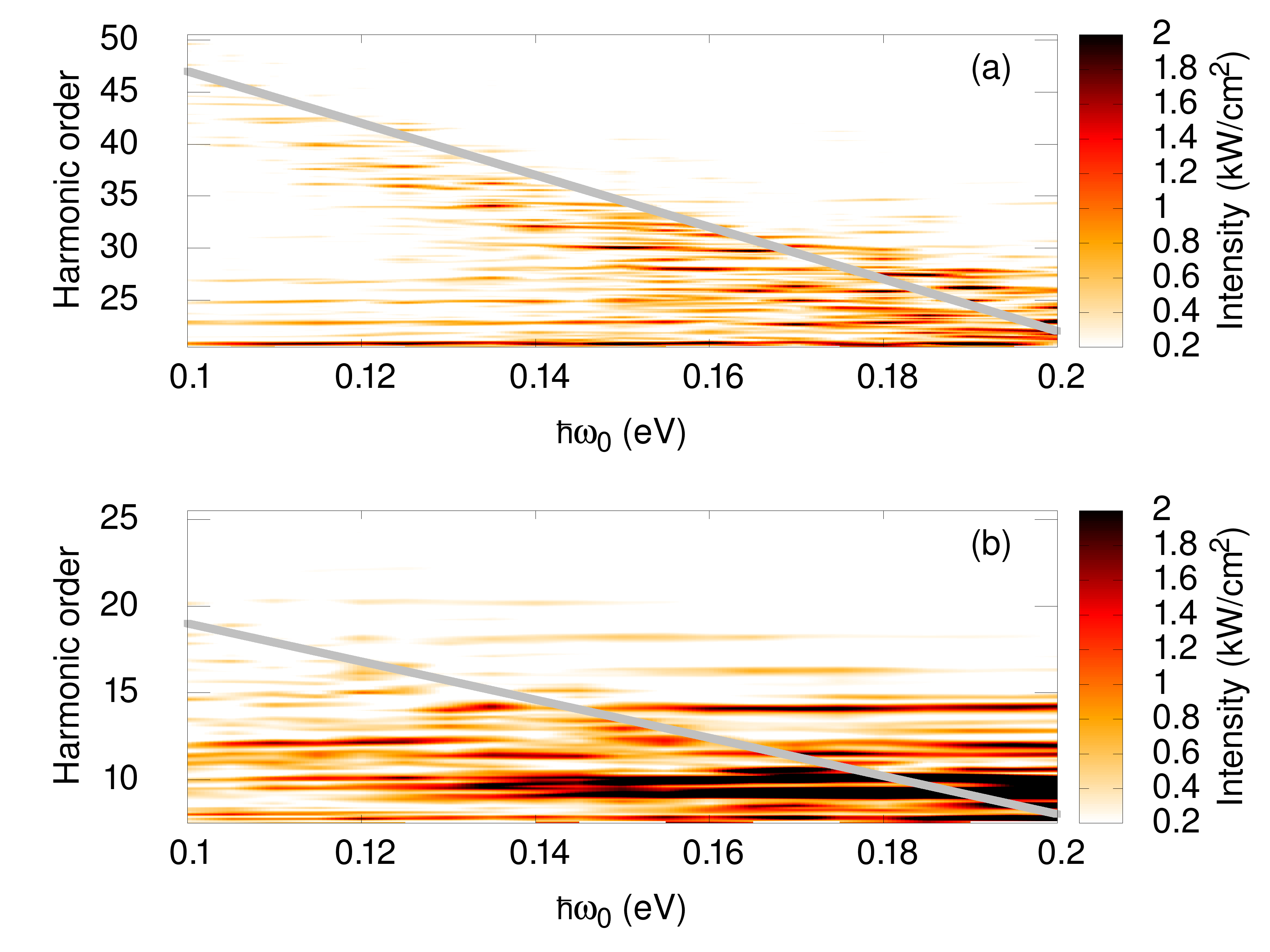}
\caption{The intensity of HHG as a function of the fundamental frequency and
the order of harmonics for graphene (a) and silicene (b) in the strong-field
regime in the bichromatic driving field: $\protect\omega _{0}^{\prime }=2%
\protect\omega _{0}$, $\hat{\mathbf{e}}=\left\{ 1,0\right\} $, $\hat{\mathbf{%
e}}^{\prime }=\left\{ 0,1\right\} $, $\protect\varphi =0$, and $\protect%
\varepsilon =1$ The interaction parameter is $\protect\chi _{0}=1.7$ and the
relaxation time is taken to be $\Gamma ^{-1}=40\ \mathrm{fs.}$ The grey
straight lines in both cases are the saddle-point excitonic resonant photon
energy.}
\end{figure}

The fingerprint of the saddle-point excitons is preserved also for the
higher intensity of laser pulses. This is seen in Fig. 10, where we plotted
the intensity of HHG as a function of the interaction parameter and
harmonic's photon energy for graphene and silicene. For comparison we also
plotted the results obtained in the scope of FP model (a) and (c). Comparing
the FP with HF approximation results we see that in the former case the
slight enhancement of HHG intensity takes place near VHS $2\gamma _{0}$,
while in the latter case the sharp enhancement for the wide rang of
intensities takes place close to the saddle-point excitonic resonances. For
graphene, this is $4-5.0\ \mathrm{eV}$ and for silicene $1.5-2.0\ \mathrm{eV}
$. This tendency is also preserved for other frequencies of the driving
field. In Fig. 11, the intensity of HHG as a function of the fundamental
frequency and the order of harmonics for graphene and silicene in the
bichromatic driving field is shown. On the same figure, we also plot the
saddle-point excitonic resonant photon energy: the harmonic order for every
driving fundamental frequency. As is seen, the sharp enhancement takes place
along the excitonic resonances.

\section{Conclusion}

We have presented the microscopic theory of nonlinear interaction of a
monolayer graphene/silicene with a strong infrared laser field near the VHS.
We have numerically solved the Bloch equations within the Houston basis that
takes into account the many-body Coulomb interaction in the HF
approximation. As reference nanostructures, we have considered graphene and
silicene. The obtained results show that saddle-point excitonic resonances
have a significant impact on the HHG process in hexagonal 2D nanostructures.
We have shown that in the bichromatic driving pulses that is composed of the
fundamental wave of linear polarization and its second harmonic at the
orthogonal polarization, one can effectively initiate spectral caustics in
the HHG spectrum. In particular, we have shown that the plateau of the HHG
spectrum has a peak near the harmonics close to SPER and redshifted from the
VHS of the free-particle picture. The results of the current investigation
are not only of theoretical/academic importance but also will have
significant implications for the rapidly developing area of modern extreme
nonlinear optics of nanostructures.

\begin{acknowledgments}
The work was supported by the Science Committee of Republic of
Armenia, project No. 21AG-1C014.
\end{acknowledgments}

\end{document}